\documentclass[
reprint,
superscriptaddress,
showpacs,preprintnumbers,
footinbib,
bibnotes,
amsmath,amssymb,
aps,
prl,
floatfix,
]{revtex4-1}

\usepackage{graphicx}
\usepackage{color}

\begin{document}


\title{Multiple photodetachment of carbon anions via single and double core-hole creation}

 \author{A.~Perry-Sassmannshausen}
 \affiliation{I.~Physikalisches Institut, Justus-Liebig-Universit\"{a}t Gie{\ss}en, Heinrich-Buff-Ring 16, 35392 Gie{\ss}en, Germany}

 \author{T.~Buhr}
 \affiliation{I.~Physikalisches Institut, Justus-Liebig-Universit\"{a}t Gie{\ss}en, Heinrich-Buff-Ring 16, 35392 Gie{\ss}en, Germany}

 \author{A.~Borovik Jr.}
 \affiliation{I.~Physikalisches Institut, Justus-Liebig-Universit\"{a}t Gie{\ss}en, Heinrich-Buff-Ring 16, 35392 Gie{\ss}en, Germany}


 \author{M.~Martins}
 \affiliation{Institut f\"{u}r Experimentalphysik, Universit\"{a}t Hamburg, Luruper Chaussee 149, 22761 Hamburg, Germany}

 \author{S.~Reinwardt}
 \affiliation{Institut f\"{u}r Experimentalphysik, Universit\"{a}t Hamburg, Luruper Chaussee 149, 22761 Hamburg, Germany}

 \author{S.~Ricz}
 \affiliation{Institute for Nuclear Research of the Hungarian Academy of Sciences, Debrecen, P.O. Box 51, 4001, Hungary}

 \author{S.~O.~Stock}
 \affiliation{Helmholtz-Institut Jena, Fr{\"o}belstieg 3, 07743 Jena, Germany}
 \affiliation{Theoretisch-Physikalisches Institut, Friedrich-Schiller-Universit\"{a}t Jena, 07743 Jena, Germany}

 \author{F.~Trinter}
 \affiliation{FS-PETRA-S, Deutsches Elektronen-Synchrotron (DESY),  Notkestra{\ss}e 85, 22607 Hamburg, Germany}
 \affiliation{Molecular Physics, Fritz-Haber-Institut der Max-Planck-Gesellschaft, Faradayweg 4-6, 14195 Berlin, Germany}

 \author{A.~M\"uller}
 \affiliation{Institut f\"{u}r Atom- und Molek\"{u}lphysik, Justus-Liebig-Universit\"{a}t Gie{\ss}en, Leihgesterner Weg 217, 35392 Gie{\ss}en, Germany}

 \author{S.~Fritzsche}
 \affiliation{Helmholtz-Institut Jena, Fr{\"o}belstieg 3, 07743 Jena, Germany}
 \affiliation{Theoretisch-Physikalisches Institut, Friedrich-Schiller-Universit\"{a}t Jena, 07743 Jena, Germany}

 \author{S.~Schippers} \email{stefan.schippers@physik.uni-giessen.de}
 \affiliation{I.~Physikalisches Institut, Justus-Liebig-Universit\"{a}t Gie{\ss}en, Heinrich-Buff-Ring 16, 35392 Gie{\ss}en, Germany}
\date{\today}

\begin{abstract}
We report on new measurements of $m$-fold photodetachment ($m=2-5$) of carbon anions via $K$-shell excitation and ionization. The experiments were carried out employing the photon-ion merged-beams technique at a synchrotron light source. While previous measurements were restricted to double detachment ($m=2$) and to just the lowest-energy $K$-shell resonance at about 282~eV, our absolute experimental $m$-fold detachment cross sections at photon energies of up to 1000 eV exhibit a wealth of new thresholds and resonances. We tentatively identify these features with the aid of detailed atomic-structure calculations. In particular, we find unambiguous evidence for fivefold detachment via double $K$-hole production.
\end{abstract}

\maketitle

Atomic anions are highly correlated systems where the extra electron is weakly bound to an overall neutral charge distribution. Consequently, the number of excited states of atomic anions is quite limited \cite{Andersen2004b} and some atomic species such as nitrogen do not form anions at all. A thorough treatment of the correlation effects in negative atomic ions still poses a formidable challenge to atomic theory, and this becomes even greater for inner-shell vacancies, since the valence electrons are then subject to strong many-electron relaxation effects following the creation of core holes \cite{Gorczyca2004a,Schippers2016a}. On the experimental side, core holes can be readily created by exciting or ionizing an inner-shell electron by a photon. For light ions, the core-hole state will subsequently decay via  Auger transitions such that electrons are emitted with the net effect of photoionization. For negative ions, the entire process is termed (multiple) photodetachment.

So far, photodetachment via the initial creation of a single $K$-hole has been experimentally studied only for a limited number of light anions up to F$^-$ \cite{Kjeldsen2001a,Berrah2001,Berrah2007a,Gibson2003a,Walter2006a,Schippers2016a,Mueller2018b}. For C$^-$ ions especially, previous measurements were carried out by Gibson et al.~\cite{Gibson2003a} and by Walter et al.~\cite{Walter2006a} who studied double photodetachment in a very narrow photon energy range just covering the $1s^2\,2s^2\,2p^3\;^4S \to 1s\,2s^2\,2p^4\;^4P$ resonance at about 282~eV. This scarcity of data is due to the fact that sufficiently high photon fluxes for more comprehensive studies were not available. Here, we present absolute cross sections, $\sigma_m$, for $m$-fold photodetachment of C$^-$ ions by a single photon,
\begin{equation}\label{eq:reaction}
h\nu + \textrm{C}^- \to  \textrm{C}^{(m-1)+} + m e^-,
\end{equation}
with $m=2,3,4,5$ and photon energies from below the $K$-edge up to $\sim$1000~eV.  These cross sections provide a view of the photodetachment dynamics of a highly correlated atomic system in unprecedented detail. Apart from the resonance at $\sim$282~eV, we observed a number of additional photodetachment resonances with a variety of line shapes as well as a clear signature for multiple detachment by direct \emph{double} $K$-hole production.  To better understand these observations, detailed atomic-structure calculations were performed by using the GRASP \cite{Joensson2013} and RATIP \cite{Fritzsche2012a} codes as well as the  Jena Atomic Calculator (JAC) \cite{Fritzsche2019}. To the best of our knowledge, there is only one other theoretical study of highly excited resonances in C$^-$ which, however, is focussed exclusively on high-spin states \cite{Piangos1998}.

\begin{figure*}
	\includegraphics[width=\linewidth]{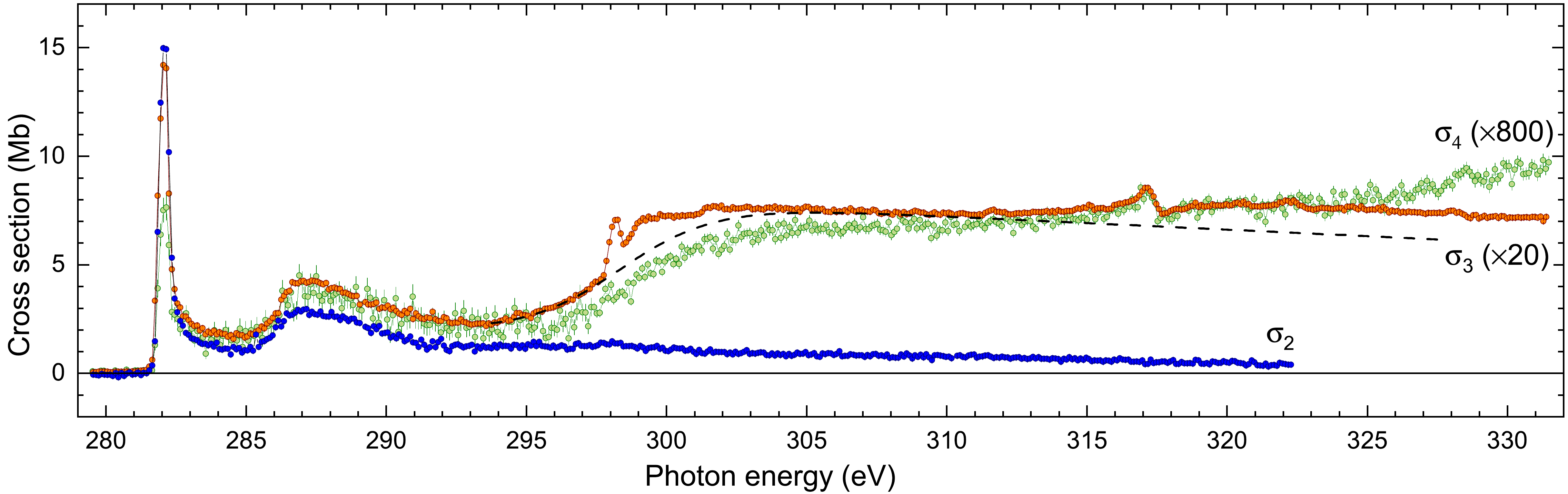}
	\caption{\label{fig:lowE}(color online) Measured cross sections for double detachment ($\sigma_2$, blue circles), triple detachment ($\sigma_3$ multiplied by a factor of 20, red circles) and fourfold detachment ($\sigma_4$ multiplied by a factor of 800, green circles) of C$^-$ ions. The dashed line represents the contribution of double detachment of a 1s and a 2p electron to $\sigma_3$ (see text). }
\end{figure*}

The experiments were carried out employing the photon-ion merged-beams technique (see \cite{Schippers2016} for a recent review) at the PIPE end-station \cite{Schippers2014} of beamline P04 \cite{Viefhaus2013} of the synchrotron radiation source PETRA\,III operated by DESY in Hamburg, Germany. Carbon anions, with practically all ions ($>99.9\%$) being in the $1s^2\,2s^2\,2p^3\;^4S_{3/2}$ ground-level \cite{Scheer1998}, were generated with a Cs-sputter ion source \cite{Middleton1983} containing solid graphite as a sputter target. After acceleration to an energy of 6~keV the ions were passed through a dipole magnet which was adjusted such that $^{12}$C$^-$ ions were selected for further transport to the photon-ion merged-beams interaction region. There, the ion beam was propagating coaxially with the soft x-ray photon beam over a distance of about 1.7~m. C$^{(m-1)+}$ ions (see Eq.~\ref{eq:reaction}) resulting from multiple photodetachment were separated from the primary ion beam by a second dipole magnet. Inside of this magnet the primary ion beam was collected in a large Faraday cup. The charge-selected product ions were passed through a spherical 180-degree out-of-plane deflector to suppress background from stray electrons, photons, and ions and then entered a single-particle detector with nearly 100\% detection efficiency \cite{Rinn1982}.

\begin{table}
\caption{\label{tab:thres} Calculated ranges of threshold energies for direct single detachment (SD), double detachment (DD), triple detachment (TD), and quadruple detachment (QD) of the C$^-$($1s^2\,2s^2\,2p^3\;^4P$) ground term as well as for SD, DD, and TD accompanied by shake-up (SU) and double shake-up (DSU) of $2s$ electrons.}
  \begin{ruledtabular}
  \begin{tabular}{lll}
    Process &  Final & Threshold \\
     &  configuration & energies (eV) \\
    \hline
    $1s$ SD & $1s\,2s^2\,2p^3$ & 282 -- 288 \\
    $1s$ SD + $2s\to 2p$ SU & $1s\,2s\,2p^4$ & 290 -- 301\\
    $1s$ SD + $2s^2\to 2p^2$ DSU & $1s\,2p^5$ & 308 -- 310\\[1ex]
    $1s+2p$ DD & $1s\,2s^2\,2p^2$ & 300 -- 305 \\
    $1s+2s$ DD & $1s\,2s\,2p^3$ & 302 -- 317 \\
    $1s+2s$ DD  + $2s\to2p$ SU & $1s\,2p^4$ & 321 -- 325\\[1ex]
    $1s+2l^2$ TD & $1s\,2l^3$ & 333 -- 356  \\[1ex]
    $1s+2l^3$ QD & $1s\,2l^2$ & 380 -- 403 \\[1ex]
    $1s^2$ DD & $\phantom{1s}\,2s^2\,2p^3$ & 657 -- 662 \\
    $1s^2$ DD + $2s\to2p$ SU & $\phantom{1s}\,2s\,2p^4$ & 666 -- 675 \\
    $1s^2$ DD + $2s^2\to2p^2$ DSU & $\phantom{1s}\,2p^5$ & 681\\[1ex]
    $1s^2+2p$ TD & $\phantom{1s}\,2s^2\,2p^2$ & 699 -- 703 \\
    $1s^2+2s$ TD & $\phantom{1s}\,2s\,2p^3$ & 701 -- 714 \\
    $1s^2+2s$ TD + $1s\to 2p$ SU & $\phantom{1s}\,2p^4$ & 719 -- 723
  \end{tabular}
  \end{ruledtabular}
\end{table}

Relative cross sections for $m$-fold photodetachment were obtained by normalizing the count rates of C$^{(m-1)+}$ product ions on the primary C$^-$ ion current and on the photon flux that was measured with a calibrated photodiode. In this procedure, fluctuations of the ion beam on short time scales, which were due to unstable operating conditions of the ion source and which caused short-term variations of the geometrical overlap with the photon beam, could not fully be accounted for. This results in a nonstatistical scatter of the experimental data points.  The cross sections were put on an absolute scale by separate measurements of the geometrical overlap of the ion and photon beams using procedures that have been described in detail elsewhere \cite{Schippers2014,Mueller2017}. The ion current in the interaction region was up to 45~nA and the photon flux  $\sim3\times10^{13}$~s$^{-1}$ at a photon energy spread of about 0.5~eV. The systematic uncertainty of the measured cross sections is estimated to be $\pm$15\% at 90\% confidence level \cite{Schippers2014}. The photon energy scale was calibrated by absorption measurements in a gas cell as described in detail in a recent publication on photoionization of C$^{4+}$ ions \cite{Mueller2018c}. The present uncertainty of the photon energy scale amounts to $\pm 0.2$~eV.

Figure \ref{fig:lowE} displays our measured cross sections for double, triple, and fourfold detachment in the energy range 279--332~eV, which comprises the threshold for the direct detachment of a single $K$-shell electron at about 281.6~eV~\cite{Walter2006a}. The most prominent feature in all three spectra is the $1s\,2s^2\,2p^4\;^4P$ resonance at 282.085~eV which has already been observed previously, but only in the double-detachment channel \cite{Gibson2003a,Walter2006a}. Towards higher photon energies, a broad asymmetric resonance feature occurs at 287~eV and is visible in all channels. Our calculations suggest that it is associated with a $1s\to 3p$ excitation to $1s\,2s^2\,2p^3\,3p\;^4P_J$ levels. Neither this feature nor any of the resonances at higher energies have been observed in the previous double-detachment experiments since it occurs outside the narrow energy ranges investigated earlier \cite{Gibson2003a,Walter2006a}.

The ratio of strengths of the 287~eV peak and the $1s\,2s^2\,2p^4\;^4P$ resonance increases with increasing product-ion charge state. The narrower $1s\,2s^2\,2p^4\;^4P$ resonance autoionizes predominantly to the lowest term ($^5S$) of the $1s\,2s^2\,2p^3$ configuration \cite{Gibson2003a}. The much larger width of the  $1s\,2s^2\,2p^3\,3p\;^4P$ resonance might be explained by the opening of additional Auger decay channels such as the $1s\,2s^2\,2p^3\;^3D$ and $^3S$ terms and possibly (depending on the precise values of the energies) even the $1s\,2s^2\,2p^3\;^3P$ and $^1D$ terms of the neutral carbon atom.

\begin{figure}
	\includegraphics[width=\linewidth]{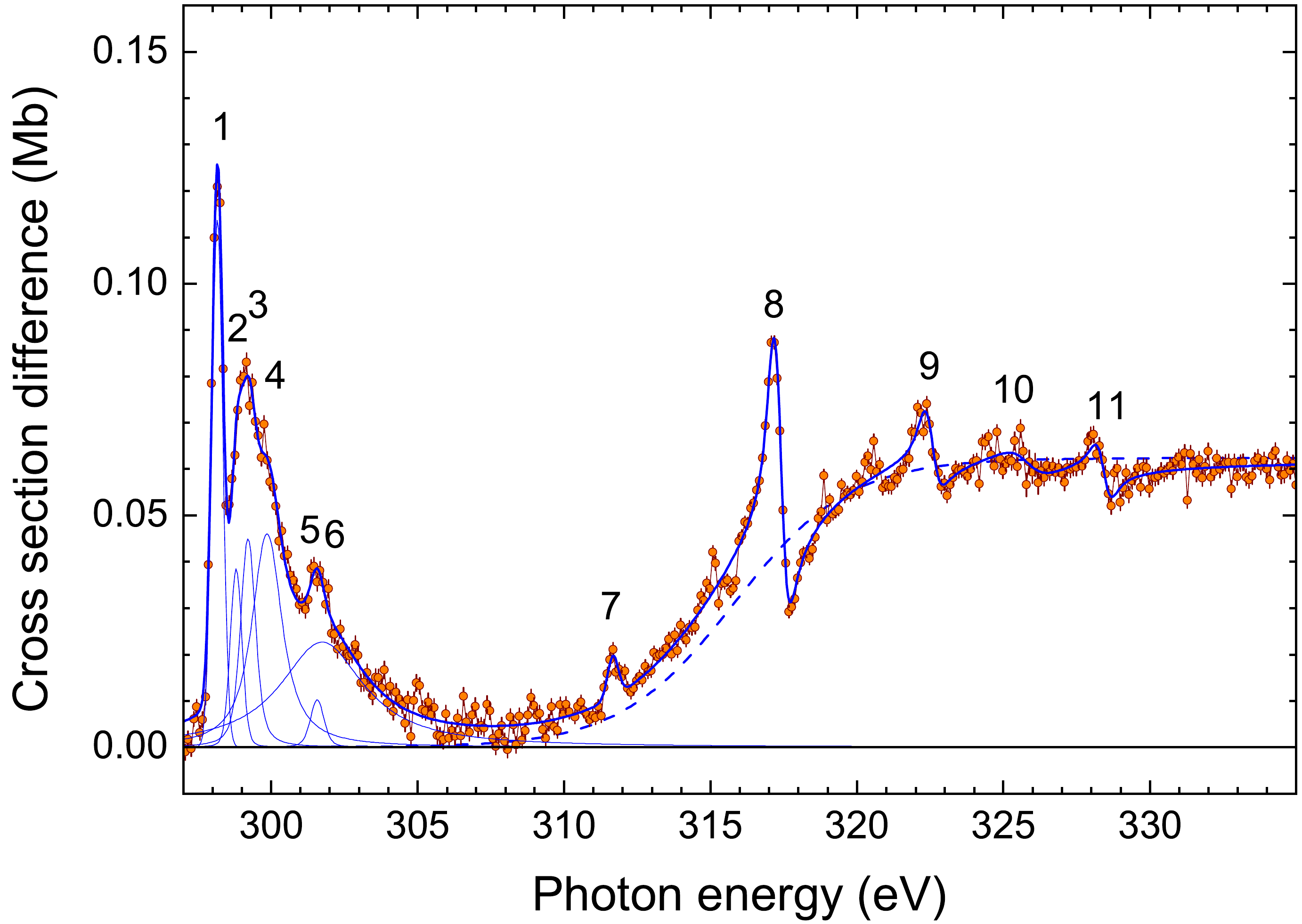}
	\caption{\label{fig:sig3d}(color online)  Cross-section difference between the measured cross section $\sigma_3$ for triple detachment of C$^-$ and the dashed line in Fig.~\ref{fig:lowE}.  The thick solid line results from a resonance fit to the experimental data. In this fit, the numbered resonance features were represented by Lorentzian and Fano line shapes convolved with a Gaussian that accounts for the experimental photon energy spread \cite{Schippers2018}. In addition, the rise of the cross section between 310 and 320~eV was taken into account by adding a smooth functional dependence (dashed line). Resonances 1--6 are shown individually as thin solid lines. Table~\ref{tab:fit} contains the resonance parameters that were obtained from this fit.}
\end{figure}

At higher energies, the different cross sections exhibit distinctively different dependences on the photon energy. The double-detachment cross section remains nearly constant up to a weak local maximum at about 298~eV, above which an almost linear decrease sets in. In contrast, the triple and fourfold detachment cross sections steeply rise between 295 and 305~eV. Our computations suggest that this rise can be attributed to the onset of the simultaneous detachment of a $1s$ electron and the shake-up of a $2s$ electron and to direct double detachment of a $1s$ and a $2p$ electron followed by the ejection of up to three Auger electrons. The calculated threshold energies for these processes occur in the photon energy ranges  290--301~eV and 300--305~eV, respectively (Table~\ref{tab:thres}). Direct $1s+2p$ double detachment has already been observed in triple-detachment of F$^-$ ions \cite{Mueller2018b}. No photodetachment resonances were observed for F$^-$ which has been attributed to its noble-gas configuration.  For C$^-$, the open $2p$ shell apparently supports a number of photodetachment resonances which are superimposed on the rising slope of $\sigma_3$. These resonances do not significantly contribute to $\sigma_4$. More resonances are visible in both $\sigma_3$ and $\sigma_4$ above 310~eV, with a particularly pronounced Fano resonance at $\sim$317.3~eV.

\begin{figure}
\includegraphics[width=\linewidth]{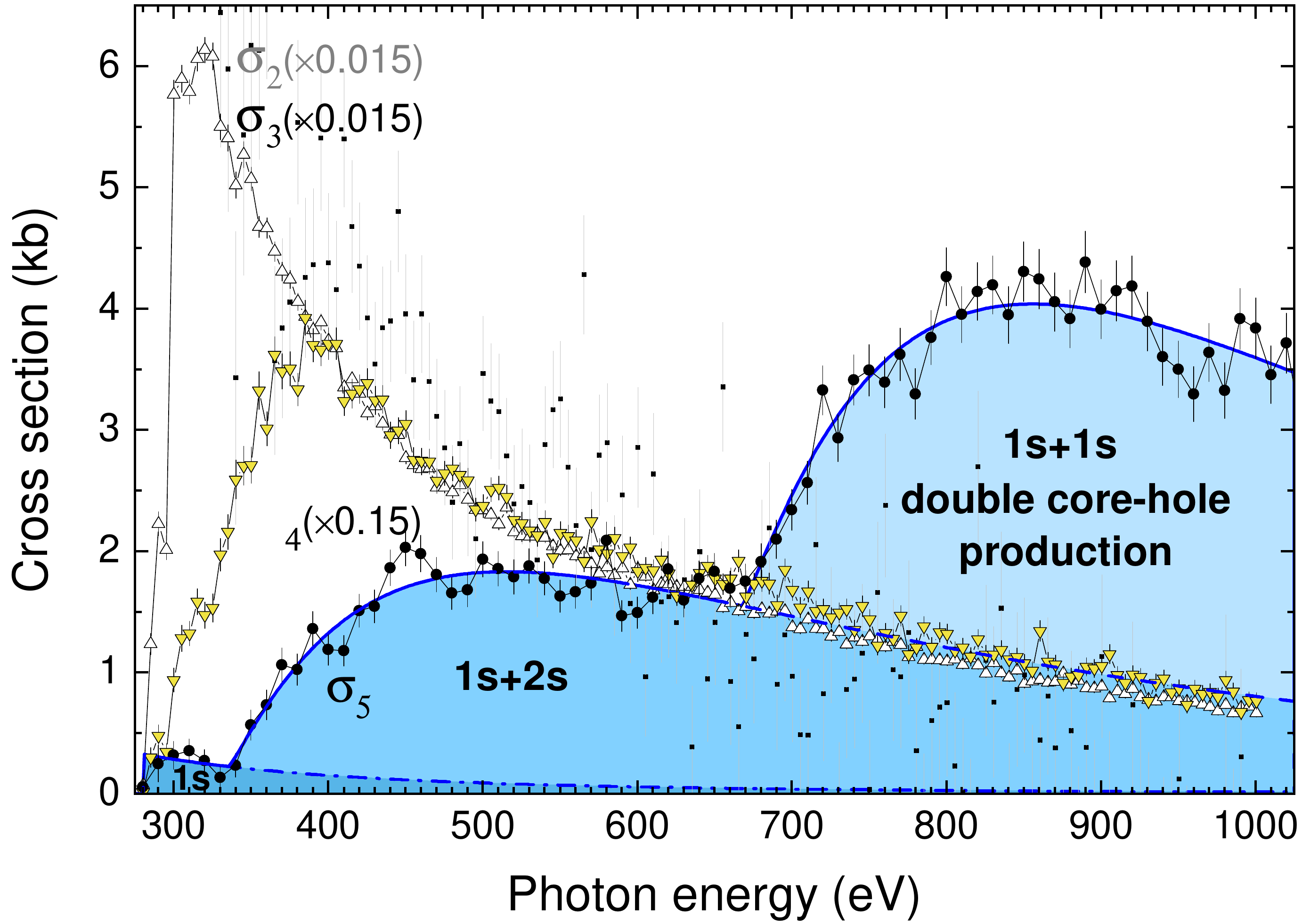}
\caption{\label{fig:all}(color online) Measured cross sections for double detachment ($\sigma_2$ multiplied by a factor of 0.015, small squares), triple detachment ($\sigma_3$ multiplied by a factor of 0.015, open triangles up), fourfold detachment ($\sigma_4$ multiplied by a factor of 0.15, triangles down), and  for fivefold detachment ($\sigma_5$, filled circles) of C$^-$ ions. $\sigma_2$ could only be measured with rather large statistical uncertainties due to the presence of a strong background in the C$^+$ product channel that was caused by collisions of the primary C$^-$ ions with residual gas particles. The shaded curves are fitted model cross sections (see text) accounting for direct single and double detachment. The dash dotted line represents the model cross section for $1s$ single detachment. The dashed line is the sum of the cross sections for $1s$ and  $1s+2s$ detachment. The full line additionally includes $1s+1s$ double K-shell detachment.}
\end{figure}

Figure~\ref{fig:sig3d} provides a close view of the contributions of these resonances to the cross section $\sigma_3$. The experimental curve in this figure was obtained by subtracting from $\sigma_3$ a smooth curve (dashed line in Fig.~\ref{fig:lowE}) that models the rise of $\sigma_3$ in the energy range 295--305~eV and its slight decrease at higher energies \footnote{The energy dependence of this curve in the energy interval 295--305~eV was obtained from a fit of an arcus tangent function to $\sigma_4$ which was subsequently added to a linear function and scaled to $\sigma_3$}. In addition to at least eleven resonance features (Table~\ref{tab:fit}), Fig.~\ref{fig:sig3d} exhibits a contribution that sets in at about 310~eV which we tentatively assign to direct $1s+2s$ double-detachment processes partly accompanied by shake-up (Table~\ref{tab:thres}). Apparently, resonant detachment above this threshold interferes with the $1s+2s$ direct double detachment leading to the observed strongly asymmetric Fano line shapes of resonances 8--11 above 316 eV. This implies that these resonances are embedded in the $1s\,2s\,2p^3\,\varepsilon l\,\varepsilon'l'$ continuum. According to our calculations, the strongest asymmetric resonance at 317.3~eV might be assigned to the $1s\,2s(^3S)\,2p^3(^4P)\,3s(^3P)\,3p\;^4P$ term.

Figure \ref{fig:all} shows the experimental cross sections for double, triple, fourfold and fivefold detachment over a much extended energy range up to $\sim$1000~eV. Here, the cross sections $\sigma_2$, $\sigma_3$ and $\sigma_4$ have been multiplied by the specified factors to scale them to the cross-section axis of the figure. For $m\geq 3$, the  cross sections $\sigma_m$ decrease roughly by a factor of 10 when $m$ is increased by 1. At photon energies larger than 400 eV, $\sigma_3$, and $\sigma_4$ show identical shapes, but differ by a factor of 10. At these energies, $\sigma_2$ has roughly the same magnitude as $\sigma_3$, but decreases slightly more steeply with increasing energy.

Figure~\ref{fig:all} also displays the cross section $\sigma_5$ for fivefold detachment which exhibits a significantly different behavior than $\sigma_2$, $\sigma_3$, and $\sigma_4$. In particular, an additional threshold at $\sim$670~eV arises, beyond which $\sigma_5$ reaches a maximum value of $\sim$4~kb at about 860~eV. The measurement of such a small cross section was only possible because of the high photon flux at the beamline, the excellent mass-separation capability of the PIPE setup, and the extremely low dark-count rate of the product-ion detector. At the cross section maximum, the count rate was only $\sim$1~Hz. Accordingly, the statistical uncertainties of the measured data points are rather high. Nevertheless, they allow for an identification of different contributions to $\sigma_5$.

\begin{table}
\caption{\label{tab:fit}Fit results from the fit displayed in Fig.~\ref{fig:sig3d}. Each resonance is characterized by its resonance energy $E_\mathrm{res}$, its natural width $\Gamma$, its asymmetry parameter $q$ ($q=\infty$ corresponds to a Lorentzian line shape), and its strength $S$. All resonances were convolved with a Gaussian with a full width at half maximum (FWHM) of 0.436(19)~eV. The resonances appear on top of a smooth cross section (dashed line in Fig.~\ref{fig:sig3d}) that was fitted as $0.03119(26)\textrm{~Mb}\,\times \{1+\mathrm{atan}[(E_\mathrm{ph}-315.488(90)\textrm{~eV})/3.89(13)\textrm{~eV}]\}$ where $E_\mathrm{ph}$ denotes the photon energy. The numbers in parentheses are the statistical uncertainties resulting from the resonance fit. The systematic uncertainty of the photon energy scale is $\pm0.2$~eV. The systematic uncertainty of the resonance strengths is $\pm15\%$.}
  \begin{ruledtabular}
  \begin{tabular}{cllcl}
    \# & $E_\mathrm{res}$ (eV)  & $\Gamma$ (eV) &  $q$ & $S$ (kb eV) \\
    \hline
1  & 298.1631(89) & $\sim$0.0 & $\infty$  & 52.9(34) \\
2  & 298.80(13)   & 0.07(46)  & $\infty$  & 21(38)  \\
3  & 299.21(16)   & 0.28(75)  & $\infty$  & 35(76)   \\
4  & 299.86(21)   & 1.06(57)  & $\infty$  & 84(76) \\
5  & 301.571(64)  & 0.19(48)  & $\infty$  & 6.9(85) \\
6  & 301.74(48)   & 3.41(41)  & $\infty$  & 122(46) \\
7  & 311.666(41)  & $\sim$0.0  & $\infty$  & 4.16(70) \\
8  & 317.335(14)  & 0.151(36) & $-$1.774(98) & 17.2(16) \\
9  & 322.504(60)  & 0.25(12)  & $-$1.77(38) & 5.7(20) \\
10 & 325.79(38)   & 1.45(70)  & $-$1.01(48) & 0.12(57) \\
11 & 328.421(94)  & 0.18(20)  & $-$0.99(33) & 0.0(20) \\
  \end{tabular}
  \end{ruledtabular}
\end{table}

The full line in Fig.~\ref{fig:all} is a model cross section that accounts for $1s$ (single) detachment as well as for $1s+2s$ and $1s+1s$ double-detachment processes. Within the experimental uncertainties, this model cross section describes the behavior of $\sigma_5$ surprisingly well. In the model, the theoretical $1s$ ionization cross section (dash-dotted line) for neutral carbon from Verner et al.\ \cite{Verner1993a} has been used for the single-detachment contribution with a scaling factor of 0.00034(7) such that the theoretical cross section matches $\sigma_5$ at energies below 340~eV. The cross sections for $1s+2s$ and $1s+1s$ double detachment were obtained from a scaling formula for multiple photoionization cross sections proposed by Pattard \cite{Pattard2002} which has recently been shown to be capable of describing double detachment of F$^-$ ions \cite{Mueller2018b}. The scaling formula is based on Wannier theory \cite{Wannier1953} and requires a threshold energy and a Wannier exponent as input parameters as well as the position and the cross-section value of the cross-section maximum. In Fig.~\ref{fig:all}, threshold energies for $1s+2s$ and $1s+1s$ double detachment of 336(4) and 668(4) eV, respectively, were determined by a fit (full line) of the model cross section to $\sigma_5$.  According to the fit, the corresponding cross-section maxima occur at photon energies of 524(10) and 891(12)~eV and amount to 1.75(4) and 3.00(9)~kb. In the fit, a Wannier exponent of 1.1269 was applied in both cases. This value corresponds to a charge of +1 of the doubly detached intermediate C$^+$ ion \cite{Wannier1953}.

The model threshold energies for $1s+2s$ and $1s+1s$ double detachment are slightly higher than the calculated threshold energies from Table~\ref{tab:thres} for these processes. We attribute this to the fact that the sudden change of the atomic potential following double detachment gives rise to a sizeable shake-up of $L$-shell electrons as has explicitly been shown, e.g., by Auger electron and photoelectron spectroscopy of double core-hole levels of rare-gas atoms \cite{Puettner2015,Southworth2003a,Goldsztejn2016}. The present structure calculations suggest that shake-up processes can shift the double-detachment thresholds of C$^-$ to higher energies by up to 20 eV. In principle, a $1s+1s+2s$ triple detachment might become possible at the calculated threshold energies in the range 699--714~eV (Table~\ref{tab:thres}) and might contribute to $\sigma_5$. This then implies a slightly lower Wannier exponent of 1.0559 \cite{Wannier1953} in the Pattard scaling formula for multiple ionization \cite{Pattard2002}. Unfortunately, our experimental data for $\sigma_5$ with their limited statistical accuracy do not allow for a discrimination of such a small change of the Wannier exponent.

Independent of a more detailed assignment, the clear rise of the cross section $\sigma_5$ above 670 eV is an unambiguous signature for double $K$-hole production by the impact of a single energetic photon. The appearance of the double $K$-hole feature in $\sigma_5$ requires that the subsequent Auger processes lead to the emission of at least three further electrons. In principle, the emission of even four electrons is energetically possible leaving a C$^{5+}$($1s$) ion behind.  However, a measurement of the cross section $\sigma_6$ for sixfold detachment has not been attempted. There is no evidence for a sizeable contribution of double $K$-hole production to any of the other experimental cross sections. In fact, $\sigma_2$, $\sigma_3$, and $\sigma_4$  follow roughly the cross-section dependence of direct $1s+2s$ double detachment also beyond the threshold for double $K$-hole creation. The slightly steeper decrease of $\sigma_2$ is easily explained by the fact that this cross section is dominated by detachment via the creation of a single $K$-hole and that the cross section for single detachment decreases faster with increasing photon energy than the cross section for double detachment.

In summary, our present experimental results on multiple detachment of carbon anions go much beyond what has been experimentally possible before. They thus provide an unprecedentedly detailed view on the complex dynamics in a highly correlated atomic system that sets in upon the creation of one or two $K$-shell holes. Our accompanying theoretical calculations allow for the identification of part of the observed thresholds and resonances. A more detailed assignment of all measured cross-section features and a quantitative treatment of the de-excitation cascades require a much larger theoretical effort including the further development of the presently available computational tools.

Our quantitative experimental results on $1s+1s$ double detachment complement previous works on double core-hole levels in atoms \cite{Southworth2003a,Goldsztejn2016,Cryan2010,Tamasaku2013}. They are also important for assessing the intensity-induced x-ray transparency of gaseous matter as observed in experiments with free-electron lasers \cite{Young2010} and the double core-hole creation in carbon-containing molecules \cite{Eland2010,Berrah2011a,Nakano2013a,Marchenko2017,Feifel2017}. Moreover, the newly discovered absorption thresholds and resonances can potentially be useful for identifying C$^-$ by x-ray absorption spectroscopy in the interstellar medium where, so far, only anions of carbon compounds have been detected \cite{Millar2017}.

\begin{acknowledgments}
We acknowledge DESY (Hamburg, Germany), a member of the Helmholtz Association HGF, for the provision of experimental facilities. Parts of this research were carried out at PETRA\,III and we would like to thank Kai Bagschik,  Frank Scholz, Jörn Seltmann, and Moritz Hoesch for assistance in using beamline P04. We are grateful for support from Bundesministerium f{\"u}r Bildung und Forschung within the \lq\lq{}Verbundforschung\rq\rq\ funding scheme (Grant Nos.\ 05K16GUC, 05K16RG1, 05K16SJA, 05K19GU3, and 05K19RG3) and from Deutsche Forschungsgemeinschaft (DFG, Project No.\ Schi~378/12-1). M.M.\ acknowledges support by DFG through project SFB925/A3.
\end{acknowledgments}


%
\end{document}